# Business Cycles explained by Instability

Galiya Klinkova, Michael Grabinski

Department of Business and Economics, Neu-Ulm University, Wileystr. 1, 89231 Neu-Ulm, Germany
Email: galiya.klinkova@uni-neu-ulm.de, michael.grabinski@uni-neu-ulm.de

## Abstract

Business cycles (a periodic change of e.g. GDP over five to ten years) exist, but a proper explanation for it is still lacking. Here we extend the well-known NAIRU (non-accelerating inflation rate of unemployment) model, resulting in a set of differential equations. However, the solution is marginal stable. Therefore we find a *natural* sinusoidal oscillation of inflation and unemployment just as observed in business cycles.
When speculation is present, the instability becomes more severe.
So we present for the first time a mathematical explanation for business cycles. The steering of central banks by setting interest rates to keep inflation stable and low needs an overhaul. One has to distinguish between real monetary instability and the one caused naturally by business cycles.

## 1. Introduction

Economic data like growth, unemployment, etc. show typically a periodic behavior. It is clear that there is a periodicity within a year. E.g. construction work is booming during summer (at least in the northern hemisphere). Weather phenomena like El Niño may give rise to a periodicity of several years. But all this is not meant by business cycles.

In a quite old textbook (Schumpeter, 1939) business cycles are *described*. However, there is no *explanation* of it. There are much newer works about business cycles such as Vianna, 2023 which is almost a review article with many recommended references. Nevertheless, it does not explain business cycles. Sometimes demography is scrutinized to give a connection to business cycles (Anderson, Botman & Hunt, 2014;

Juselius & Taktis, 2015; Katagiri, Konishi & Ueda, 2020; The Economist, 2015). But again, it is far from an explanation.

The Economist explicitly states "Economists still lack a proper understanding of business cycles" (Economist, 2018). Even Covid 19 may (or may not) affect business cycles (Economist, 2024).

Here we will give rigorous *explanation* for business cycles for the first time. Stability (see e.g. Bronshtein et. al. 2007) is relative rarely scrutinized in economic models. In most cases equilibrium states are calculated only. In Schädler & Grabinski, 2015 the NAIRU (Non-Accelerating Inflation Rate of Unemployment) model is extended in order to have a set of soluble differential equations. Besides its solution stability has been analyzed in Schädler & Grabinski, 2015. Depending on the parameters, the extended NAIRU model shows typically a marginal instability which automatically leads to an oscillation with a period of several years.

The rest of the paper is organized as follows. Chapter 2 summarizes Schädler & Grabinski, 2015 as far as it is necessary here. In chapter 3 this is applied to business cycles. In chapter 4 we close with conclusions and future work.

## 2. The extended NAIRU Model

Besides comparative advantage, NAIRU is one of the very few widely accepted models in economics. The logic behind it is as follows. When the unemployment $u(t)$ becomes big, labor becomes cheaper. As salaries for the same work decrease, one has deflation or decreased inflation $I(t)$. When unemployment becomes small, labor costs will increase (and with-it inflation) because there is a possible shortage of workers. Please note that unemployment is always positive ($u(t) \geq 0$). So small unemployment means $u(t) < n$ where $n$ is an equilibrium unemployment. Putting this together leads to the well-known formula for NAIRU:

$$\partial_t I(t) = -a \cdot (u(t) - n) \quad with \quad a, n > 0 \qquad (1)$$

In order to have stable inflation $\partial_t I(t) = 0$ or $u(t) = n$. This is the stationary solution of the differential Equation (1). Please note that a stationary solution does not say anything about how fast or slow an equilibrium is reached. And for sure it does not tell anything about the stability. The differential equation (1) is one equation for two functions $I(t)$ and $u(t)$. The solution is therefore undetermined. The necessary second equation is

$$\partial_t u(t) = \frac{\kappa}{a} \cdot (I(t) - z) \cdot (1 - u(t)). \qquad (2)$$

A rigorous derivation of Equation (2) can be found in Schädler & Grabinski, 2015. $z$ denotes the equilibrium inflation and $\kappa > 0$ is a new constant. Equations (1) and (2) are a set of non-linear differential equations. Equations (1) and (2) can be solved numerically only (though very easily). In order to scrutinize stability, the differential equations must be linearized around their equilibrium:

$$I(t) \equiv z + \varepsilon(t) \quad and \quad u(t) \equiv n + \eta(t) \qquad (3)$$

Inserting this into Equations (1) and (2) leads to two differential equations:

$$\partial_t \varepsilon(t) = -a \cdot \eta(t) \tag{4}$$

$$\partial_t \eta(t) = \frac{\kappa}{a} \cdot \varepsilon(t) \cdot (1 - n - \eta(t)) \tag{5}$$

Equations (4) and (5) are mathematically identical to Equations (1) and (2). They can be decoupled to:

$$\ddot{\varepsilon} = -\kappa \cdot \varepsilon \cdot \left(1 - n + \frac{1}{a} \cdot \dot{\varepsilon}\right) \tag{6}$$

$$\ddot{\eta} = -\kappa \cdot \eta \cdot (1 - n + \eta) - \frac{\dot{\eta}^2}{1 - n - \eta} \tag{7}$$

All these equations can be found also in Schädler & Grabinski, 2015. Please note that there is a typo in the equation for $\ddot{\eta}$ in Schädler & Grabinski, 2015 (also Equation (7) there). By skipping the parts marked red in Equations (6) and (7), we get a linearized version. (These can be solved easily. Except for triple-digit inflation, their solution is hardly different from the non-linear ones)

The linearized version of Equations (4) and (5) can also be written as:

$$\partial_t \begin{pmatrix} \varepsilon(t) \\ \eta(t) \end{pmatrix} = \begin{pmatrix} 0 & -a \\ \frac{\kappa}{a} \cdot (1 - n) & 0 \end{pmatrix} \begin{pmatrix} \varepsilon(t) \\ \eta(t) \end{pmatrix} \tag{8}$$

This is the standard form to scrutinize stability. The matrix in Equation (8) has the eigenvalues

$$\lambda_{1,2} = \pm\sqrt{-\kappa \cdot (1 - n)}. \tag{9}$$

For $\text{Re}(\lambda) < 0$ we would have stability. $\text{Re}(\lambda) > 0$ would mean instability. As $\kappa > 0$ and the equilibrium unemployment $n$ is above zero and below 1 (100% unemployment), we see that $\lambda$ is purely imaginary. In other words, we have marginal stability.

In Schädler & Grabinski, 2015 it is also considered that the capital comes from realwirtschaft *and* speculation which is always the case in the real world. Under most circumstances it will lead to $\text{Re}(\lambda) > 0$. So we have instability which makes setting interest rates by central banks essentially ludicrous. But this is left to further research.

Here we will stick to Equation (9) which will easily imply business circles as observed and considered in the next chapter.

## 3. Business Cycles derived from Instability

Equations (1) and (2) are a simple model to describe the connection between unemployment $u(t)$ and inflation $I(t)$. As central banks have the job to keep inflation at bay, they adjust the interest rate $r \gtrless z$ in order to control inflation and with it (whether intended or not) unemployment $u(t)$. When inflation is high ($I(t) > z$), they will raise the interest rate in order to curb inflation. As the real world is slightly more complicated, so is the work of central banks. One obstacle is that the constants $n$ and $z$ are not given naturally. They may vary from economy to economy and in the long run also in time. However, all these problems are negligible compared to the problem stated here. Of course one solution of Equations (1) and (2) is:

$$I(t) \equiv z \quad \text{and} \quad u(t) \equiv n \tag{10}$$

Central bankers might say that the status in Equation (10) is the goal they are trying to achieve. However, as inflation and unemployment will never *exactly* become $z$ and $n$, respectively, the solution in Equation (10) will *never* be given. One has to solve Equations (6) and (7) in order to get the true connection between inflation and unemployment. The non-linear parts in Equations (6) and (7) can be neglected so unemployment $u(t)$ and inflation $I(t)$ will show a natural sinusoidal behavior. The general (*quasi stationary* or *weakly stationary* as economists call it) solution of the linearized version of Equations (1) and (2) is:

$$I(t) = z + a \frac{z - I(0)}{\sqrt{\kappa - n\kappa}} \sin(t\sqrt{\kappa - n\kappa}) \quad (11)$$

$$u(t) = n - (n - u(0))\cos(t\sqrt{\kappa - n\kappa}) \quad (12)$$

In order to keep it simple we have chosen the initial conditions as $\dot{u}(0) = 0$ and $\dot{I}(0) = a(z - I(0))$. Any other initial conditions are possible too. As any set of linear differential equations can be solved by Fourier transformation, the most general solution will be like Equations (11) and (12) with some phase in the sin and cos functions. Assuming 2% inflation ($z$) and 5% unemployment ($n$), Equations (11) and (12) can be plotted as:

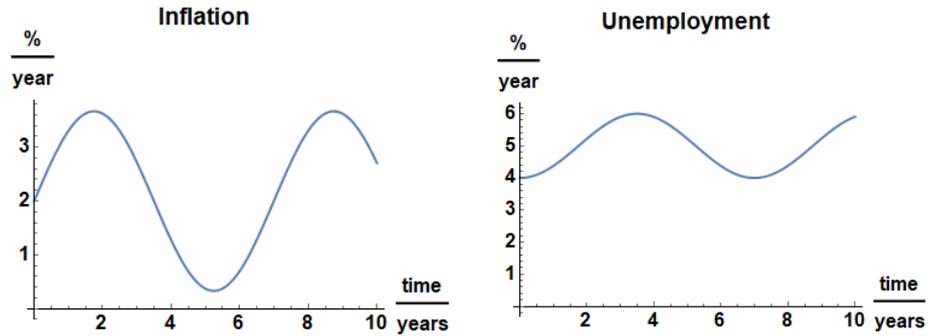

**Figure 1.** Solution of Equations (11) and (12) with $z = 2/100\text{y}, n = 5/100, a = 15/10\text{y}^2, \kappa = 85/100\text{y}^2$ where y stand for *year*.

The most striking result of Figure 1 is that the equilibrium is not given by constant inflation and unemployment like in Equation (10). It has a *natural* oscillation. **Figure 1** shows a period of roughly seven years. Unemployment is always a quarter of the period (here roughly 21 months) "behind" inflation. Raising interest rates in order to curb inflation becomes now at least doubtful as **Figure 1** shows a natural behavior.

Changing the interest rate will change the inflation accordingly. This will imply a calculation like above with changed initial conditions. Deviating from $\dot{u}(0) = 0$ and $\dot{I}(0) = a(z - I(0))$ will not change the amplitudes in **Figure 1**.

The constants in Equations (9) and (12) ($a$ and $\kappa$) are measurable quantities rather than fit parameters. In order to understand $a$ look at Equation (1). It is the speed how fast prices accelerate when unemployment is below $n$. In our plot we used $a = 15/10y^2$. When the unemployment is one percentage point below the equilibrium, it will take a year until the inflation goes up by 1.5 percentage points.

Now consider $\kappa/a$ in Equation (2). $\kappa/a$ is the factor how much unemployment changes due to a deviation of $I(t)$ from $z$. In our plot in **Figure 1** we have $\kappa/a \approx$ 0.6. So if inflation $I(t)$ is one percentage point above the equilibrium $z$, it will take a year that unemployment raise by about 0.6 percentage points. As $\kappa = a \cdot \kappa/a$ we can state that $\kappa$ is the product of speed of changes in inflation due to deviation of unemployment from equilibrium times the speed of changes in unemployment due to deviations of inflation from equilibrium.

Though $a$ and $\kappa$ are measurable in principle, it does not mean that they are easy to measure in reality. The tricky points in measuring $a$ and $\kappa$ is that everything else should stay identical which is next to impossible. Furthermore, $a$ and $\kappa$ are not natural constants. Especially, they will have different values for different industries. In that sense Equations (1) and (2) will not hold for an entire economy but for each industry within an economy. As the differential Equations (1) and (2) are essentially linear as their non-linear parts are negligible, one may use the arithmetic average for $a$ and $\kappa/a$ in Equations (1) and (2). This surprising fact has been proven quite recently by Grabinski & Klinkova, 2019. The time dependence of $a$ and $\kappa$ is trickier to handle.

The frequency $\sqrt{\kappa - n\kappa} = \sqrt{\kappa(1-n)}$ in Equations (11) and (12) is essentially $\sqrt{\kappa}$. From above we know that $\sqrt{\kappa}$ is the geometric mean of the speed of unemployment $(\kappa/a)$ and inflation $(a)$ change.

## 4. Conclusions and further work

We have shown that business cycles of say five to ten years appear naturally due to an instability. Though it is the right goal to keep inflation low and constant, it appears far too short fetched to adjust interest rates in order to curb inflation.

From Equations (11) and (12) we see that the amplitudes of these oscillations are given by:

$$a \frac{z - I(0)}{\sqrt{\kappa - n\kappa}} \qquad and \qquad n - u(0) \qquad (13)$$

As $a$ and $\kappa$ are constants at a certain time in a particular economy or even industry, it appears at first glance as a panacea to try to reach $I = z$ and $u = n$ simultaneously. This is however too good to be true. Nobody knows the values of $z$ and $n$ (exactly). They will also depend on the economy or even industry. Furthermore, nobody can *set* initial conditions here.

But what should central banks do in order to keep inflation low and constant? Firstly, one has to accept that the goal of $I(t) = z$ and $u(t) = n$ (cf. Equation (10)) is impossible to reach. Secondly, one has to accept a picture like in **Figure 1**. It is then essential to at least estimate where in the business cycles the economy is. As a last step one may change the interest rate.

As already stated in Schädler & Grabinski, 2015 one has also to accept that inflation and unemployment is much trickier than thought. It is impossible to define inflation properly as an *identical* product or service does not exist. So it is useless to measure an increased or decreased price. But even less theoretical it is hard to tell whether we have a smart phone inflation or deflation. Of course, every year prices are going up. On the other hand, a smart phone with the power of a ten-year-old one will cost almost nothing today (if it is sold at all). Even during the recent gas shortage due to the Ukrainian war this fact has been ignored. In Germany energy prices were essentially subsidized by giving money to German citizens. That raised inflation (and debt). In Spain they did essentially the same but they lowered taxes on energy products. This caused (formerly) much less inflation and about the same amount of debt.

From a perspective of true inflation both were wrong. There was no inflation on e.g. gas. Russian gas is a *different* product from e.g. gas from Oman.

Unemployment is at first glance easy to define. It is clear whether somebody is working or not. Even part time jobs (if counted partly) are easy to handle. However, it is misleading to say that a ten-year-old or eighty-year-old is unemployed though both are most likely not working. Even within a family one spouse may work (outside the house) for many hours while the other one takes care for the house and the kids. So we have one employed person. If both spouses work outside the house, they may hire a nanny which will lead to say 2 ½ employed people though nothing really changed. Of course unemployment agencies will try to account for this problems. Though their rules are strict, they bear a high amount of arbitrariness.

There are two obvious extensions for this paper. One is speculation and one are non-linearities. It is straight forward to include speculation in this model. It is essentially already done in Schädler & Grabinski, 2015. The main point in speculation is a change in Equation (2). Investing the proceeds of the realwirtschaft one does not need to borrow and interest rates are almost uninteresting. One wants to have a high return (after inflation) on the invested money. Contrary to this, most speculation in banking products is done with borrowed money. So interest is the essential cost factor. The details can be found in Schädler & Grabinski, 2015. The inclusion of speculation is here absolutely straight forward; we omitted it in order to make the publication not too clumsy. Nevertheless it is important to see that the inclusion of speculation makes the situation much worse. The eigenvalue $\lambda$ is slightly more complicated than in Equation (9) and we will have partly the situation that $Re(\lambda) > 0$. So besides a sinusoidal oscillation we will have (partly) exponential growth in **Figure 1**. The effect becomes more severe if more money comes from speculation.

In that situation it can be absolutely counter productive to change the interest rate in accordance to inflation. A cut in interest rates makes borrowing cheaper. Theoretically this will push investment in banking products *and* machines, etc. However, a one percentage point change in interest will almost have no effect in the realwirtschaft as everybody knows who made calculations for major business investments. For speculation a one percentage point change in interest is tremendous. So if inflation is not too high, central bankers might lower the interest rate in order to increase GDP. And indeed one will see positive effects on the economy. However, they are due to the proceeds from speculation. And as stated above this will lead to much severe instability.

Even in the present model (Equations (1) and (2)) we have non-linearities. Including speculation the situation will not change. How about an extended model which will show more severe non-linearities? In Appendix A of Schädler & Grabinski, 2015 there is a rigorous derivation of Equation (2) and especially the version which includes speculation. From this we see that our model is the lowest order non-trivial model derived by Taylor expansion. As the non-linearities in our model does not cause any change it looks far fetched that higher orders will bring any quantitative change. Though this is no mathematical proof.

A quite other source of non-linearities will for sure have effects. Our model has no limitations. So an infinite amount of labor and other resources is assumed. This is of course far form reality. In reality one needs a limitation like e.g. in the spread of diseases. At start the number of infections grow purely exponential. As not more than everybody can be infected, there must be a limit. Including it will lead to bell-shaped curves, for details see e.g. Grabinski & Klinkova, 2020. It is clear that the limit will lead to non-linearities which cannot be neglected. Handling these non-linearities is probably very simple with numerical means. However, one cannot just take averaged values for the constants, for details please see Grabinski & Klinkova, 2019. The constants here are $a$ and $\kappa/a$. They may have different values for e.g. the automotive industry and banking and taking the average (even weighted average) is not allowed. To fix this problem will be far from being easy. One has to find a proper continuum limit.

## Conflicts of Interest

The authors declare no conflicts of interest regarding the publication of this paper.